\RequirePackage{snapshot}
\documentclass[preprint2]{aastex}
\usepackage{natbib}
\usepackage{graphicx}
\usepackage{amsmath}
\usepackage{amssymb}
\usepackage{epstopdf}
\usepackage{natbib}
\usepackage[xindy, toc, hyperfirst=false, nolist, nostyles, sanitize={name=false,description=false,symbol=false}]{glossaries}
\glsdisablehyper

\newcommand{\plotdir}{plots}

\newcommand{\msun}{\hbox{$M_{\odot}$}}

\newcommand{\lsun}{\hbox{$L_{\odot}$}}

\newglossaryentry{vrad}{name={radial velocity~}, text={radial velocity}, symbol={\ensuremath{v_\textrm{rad}}}, description={radial velocity}, sort=vrad}
\newglossaryentry{vrot}{name={stellar rotation~}, name={stellar rotation}, symbol={\ensuremath{v_\textrm{rot}}}, description={radial velocity}, sort=vrot}
\newcommand{\vrot}{\glssymbol*{vrot}}
\newcommand{\vrad}{\glssymbol*{vrad}}

\newcommand{\rasc}[4]{\hbox{\ensuremath{\alpha=#1^h#2^m#3\overset{s}{.}#4}}}
\newcommand{\decl}[3]{\hbox{\ensuremath{\delta=#1^{\circ}#2\arcmin#3\arcsec}}}
\newcommand{\kpc}{kpc}

\newcommand{\kms}{\ensuremath{\textrm{km}~\textrm{s}^{-1}}}

\newcommand{\xray}{X-ray}

\newglossaryentry{angstrom}{name=\AA, description={unit of length $10^{-10}$\,m}, sort=angstrom}
\newglossaryentry{nir}{name=NIR,description={near infrared},first = {near infrared (NIR)}}
\newglossaryentry{psf}{name=PSF,description={Point Spread Function},first = {PSF}}
\newglossaryentry{fwhm}{name=FWHM,description={Full Width Half Maximum},first = {FWHM}}
\newglossaryentry{rms}{name=RMS,description={Root Mean Square},first = {RMS}}
\newglossaryentry{signalnoise}{name=signal-to-noise,description={signal to noise}}
\newglossaryentry{uv}{name=UV,description={ultra violet},first = {ultra violet (UV)}}
\newglossaryentry{halpha}{name=\ensuremath{\textrm{H}\alpha}, description={First line of the Balmer series at 6563\,\AA}, sort=halpha}
\newglossaryentry{mgb}{name={Mg \textsc{i} b}, description={Triplet at 5167\,\AA, 5173\,\AA and 5184\,\AA}}
\newglossaryentry{sobolevapprox}{name={Sobolev approximation}, description={Lines are approximation with an infinitley thin interaction region \citep[e.g. no broadening][]{1960mes..book.....S}}, first={Sobolev approximation }}
\newglossaryentry{radeq}{name={radiative equilibrium}, description={The net flux of energy between matter and radiation field is zero}}
\newglossaryentry{nebularapprox}{name={nebular approximation}, description={Assumes that the plasma condition are controlled by a central radiation source. The radiation field decreases with the distance to the source by geometrical dilution. See \citet{1978stat.book.....M} for details}}
\newglossaryentry{modnebularapprox}{name={modified nebular approximation}, description={In contrast to \gls{nebularapprox} where only geometrical dilution is taken into account, the modified nebular approximation also takes dilution by other radiative processes into account }, first={modified nebular approximation}, parent=nebularapprox}
\newglossaryentry{thompsonscat}{name={Thomson scattering}, description={Scattering of photons on low energy electrons}}
\newglossaryentry{lte}{name={LTE}, description={Local Thermodynamic Equilibrium}, first={local thermodynamic equilibrium (LTE)}}
\newglossaryentry{lsr}{name={LSR}, description={Local Standard of Rest}, first={\textit{local standard of rest} (LSR)}}
\newglossaryentry{mc}{name={MC}, description={Monte Carlo}, first={\textit{Monte Carlo} (MC)}}
\newglossaryentry{cmf}{name=CMF, text=CMF, first=Comoving Frame (CMF henceforth)}

\newglossaryentry{sfit}{name=SFIT, text=\textsc{sfit}, description={spectral fitting program for hot stars \citep{2001A&A...376..497J}}, first={\textsc{sfit} \citep{2001A&A...376..497J}}}
\newglossaryentry{iraf}{name=IRAF, text=\textsc{iraf}, description={Image Reduction and Analysis Facility maintained by NOAO}, first={\textsc{iraf}\protect\footnote{IRAF: the Image Reduction and Analysis Facility is distributed by the National Optical Astronomy Observatory, which is operated by the Association of Universities for Research in Astronomy (AURA) under cooperative agreement with the National Science Foundation (NSF).}}}
\newglossaryentry{pyraf}{name=PyRAF, text=\textsc{PyRAF}, description={Python wrap of \gls{iraf} maintained by STSCI}, first=\textsc{PyRAF} \protect\footnote{PyRAF is a product of the Space Telescope Science Institute, which is operated by AURA for NASA.}}
\newglossaryentry{scipy}{name=SciPy, text=\textsc{Scipy}, description={Scientific Python \cite{Jones:2001fk}}}
\newglossaryentry{moog}{name=MOOG,text={\textsc{moog}}, description={spectral synthesis software \citep{1973ApJ...184..839S}}, first={\textsc{Moog} \citep{1973ApJ...184..839S}}}
\newglossaryentry{atlas9}{name=ATLAS9,description={grid of stellar atmospheres \citep{2004astro.ph..5087C}}, first={ATLAS9 \citep{2004astro.ph..5087C}}}
\newglossaryentry{vald}{name=VALD,description={Vienna Atomic Line Database \citep{2000BaltA...9..590K}}, first={Vienna Atomic Line Database \citep[VALD;][]{2000BaltA...9..590K}}}
\newglossaryentry{sextractor}{name=SExtractor, text=\textsc{SExtractor}, description={Source Extractor photometry program \citep{1996A&AS..117..393B}}, first={\textsc{SExtractor} \citep{1996A&AS..117..393B}}}
\newglossaryentry{idl}{name=IDL,text={\textsc{idl}}, description={Interactive Data Language}}
\newglossaryentry{makee}{name=MAKEE,text=\textsc{makee}, description={MAuna Kea Echelle Extraction by Tom Barlow available}}
\newglossaryentry{minuit}{name=MINUIT,text={\textsc{minuit}}, description={collection of numerical optimization tools \citep{James:1975dr}}}
\newglossaryentry{migrad}{name=MIGRAD,text={\textsc{migrad}}, description={numerical gradient optimization tools - part of \gls{minuit}}}
\newglossaryentry{dolphot}{name=DOLPHOT, text=DOLPHOT, description=photometry package for HST, first=DOLPHOT \citep{2000PASP..112.1383D}}
\newglossaryentry{chianti}{name=CHIANTI, text=CHIANTI, description= CHIANTI Database 7.1, first =CHIANTI 7.1 \citep{1997A&AS..125..149D,2012ApJ...744...99L}}
\newglossaryentry{synpp}{name=SYNPP, text=SYN++, description= SYN++ software, first =SYN++ \citep{2011PASP..123..237T}}
\newglossaryentry{tardis}{name=TARDIS, text=TARDIS, description= TARDIS MC code, first = Temporal And Radiative Diffusion In Supernovae (TARDIS)}

\newglossaryentry{2mass}{name=2MASS,description={Two Micron All Sky Survey \citep{2006AJ....131.1163S}}, first={Two Micron All Sky Survey \citep{2006AJ....131.1163S}}}
\newglossaryentry{nomad}{name=NOMAD,first={Naval Observatory Merged Astrometric Dataset \citep[NOMAD; ][]{2005yCat.1297....0Z}}}
\newglossaryentry{wifes}{name=WIFES, text=\textsc{WiFeS}, first={\textsc{WiFeS} \citep{2007Ap&SS.310..255D}},  description={Wide Field Spectrograph - \gls{ifu} mounted on the 2.3\,m telescope at Siding Spring Observatory}}
\newglossaryentry{scp}{name=SCP,description={Supernova Cosmology Project, led by Saul Perlmutter}, first={Supernova Cosmology Project (SCP)}}
\newglossaryentry{hzsns}{name=HZSNS,description={High Z Supernova Search, led by Brian Schmidt}, first={High Z Supernova Search (HZSNS)}}
\newglossaryentry{vlt}{name=VLT,description={Very Large Telescope located on Cerro Paranal (Chile)}, first={Very Large Telescope (VLT)}}
\newglossaryentry{flames}{name=FLAMES,description={Multi-object, intermediate and high resolution spectrograph mounted on the  \gls{vlt}}}
\newglossaryentry{hires}{name=HIRES, description={High Resolution Echelle Spectrometer mounted on the Keck Telescope}, first={High Resolution Echelle Spectrometer \citep[HIRES;][]{1994SPIE.2198..362V}}}
\newglossaryentry{lris}{name=LRIS,description={Low Resolution Imaging Spectrometer mounted on the Keck Telescope}, first={Low-Resolution Imaging Spectrometer \citep[LRIS;][]{Oke95}}}

\newglossaryentry{essence}{name=ESSENCE,description={The `Equation of State: SupErNovae trace Cosmic Expansion' project \citep[ESSENCE;][]{2002AAS...201.7809G}}, first={`The Equation of State: SupErNovae trace Cosmic Expansion' \citep[ESSENCE;][]{2002AAS...201.7809G}}}
\newglossaryentry{ifu}{name=IFU,description={Optical instrument combining spectrographic and imaging capabilities, used to obtain spatially resolved spectra}, first={Integral Field Unit (IFU)}, firstplural={Integral Field Units (IFUs)}} 

\newglossaryentry{besancon}{name=Besan\c{c}on Model, description={Model of stellar population synthesis of the Galaxy, including kinematics.}, first =Besan\c{c}on Model \citep{2003A&A...409..523R}}

\newglossaryentry{int}{name=INT,description={Isaac Newton 2.5\,m Telescope}, first={Isaac Newton 2.5\,m Telescope (INT)}}
\newglossaryentry{iau}{name=IAU,description={International Astronomical Union}, first={IAU}}
\newglossaryentry{chandra}{name=Chandra,description={Chandra \xray\ Observatory (space-based)}}
\newglossaryentry{hst}{name=HST,description={Hubble Space Telescope}}
\newglossaryentry{wfpc2}{name=WFPC2,description={Wide-Field Planetary Camera 2 mounted on the \gls{hst}}, first={Wide-Field Planetary Camera 2 (WFPC2)}}
\newglossaryentry{acs}{name=ACS,description={Advanced Camera for Surveys mounted on the \gls{hst}}, first={Advanced Camera for Surveys (ACS)}}
\newglossaryentry{snls}{name=SNLS,description={Supernova Legacy Survey \citep{2003AAS...203.8209P}}, first={Supernova Legacy Survey \citep[SNLS;][]{2003AAS...203.8209P}}}
\newglossaryentry{dass}{name=DASS, description={Digitized Astronomy Supernova Survey \citep{1975PASP...87..565C}}, first={Digitized Astronomy Supernova Survey \citep[DASS;][]{1975PASP...87..565C}}}
\newglossaryentry{bait}{name=BAIT, description={Berkley Automatic Imaging Telescope \citep{1993PASP..105.1164R}}, first={Berkley Automatic Imaging Telescope \citep[BAIT;][]{1993PASP..105.1164R}}}
\newglossaryentry{kait}{name=KAIT, description={Katzman Automatic Imaging Telescope \citep{2001ASPC..246..121F}}, first={Katzman Automatic Imaging Telescope \citep[KAIT;][]{2001ASPC..246..121F}}}
\newglossaryentry{loss}{name=LOSS, description={Lick Observatory Supernova Search  \citep{2000AIPC..522..103L}}, first={Lick Observatory Supernova Search \citep[LOSS;][]{2000AIPC..522..103L}}}
\newglossaryentry{ctss}{name=CTSS,description={Cal\'{a}n/Tololo Supernova Survey \citep{1993AJ....106.2392H}}, first={Cal\'{a}n/Tololo supernova survey \citep[CTSS;][]{1993AJ....106.2392H}}}
\newglossaryentry{ctio}{name= CTIO, description={Cerro Tololo Inter-American Observatory}, first={Cerro Tololo Inter-American Observatory (CTIO)}}
\newglossaryentry{ptf}{name=PTF, description={Palomar Transient Factory \citep{2009PASP..121.1334R}}, first={Palomar Transient Factory \cite[PTF;][]{2009PASP..121.1334R}}}
\newglossaryentry{batse}{name=BATSE, description={Burst and Transient Source Experiment mounted on the Compton Gamma Ray Observatory}, first={Burst and Transient Source Experiment (BATSE)}}
\newglossaryentry{bepposax}{name=BeppoSAX, description={\xray\ satellite named in honor of Giuseppe "Beppo" Occhialini}}
\newglossaryentry{rosat}{name=ROSAT, description={short for R\"{o}ntgensatellit}, first={ROSAT}}
\newglossaryentry{hete2}{name=HETE2, description={High Energy Transient Explorer}, first={High Energy Transient Explorer (HETE)}}
\newglossaryentry{gnirs}{name=GNIRS, description={Gemini Near InfraRed Spectrograph mounted on the Gemini North Telescope}}
\newglossaryentry{swift}{name=Swift, description={Swift Gamma-Ray Burst Mission}}
\newglossaryentry{vla}{name=VLA, description={Very Large Array radio telescope located in North America}, first={Very Large Array (VLA)}}
\newglossaryentry{evla}{name=EVLA, description={Extended Very Large Array radio telescope located in North America}, first={Extended Very Large Array (EVLA)}}
\newglossaryentry{sdss}{name=SDSS, description={Sloan Digital Sky Survey}}
\newglossaryentry{dss}{name=DSS, description={Digitized Sky Survey}}
\newglossaryentry{skymapper}{name=SkyMapper, description={SkyMapper telescope \citep{2007PASA...24....1K}}, first={SkyMapper \citep{2007PASA...24....1K}}}
\newglossaryentry{panstarrs}{name=PanSTARRS, description={Panoramic Survey Telescope \& Rapid Response System \citep{2004SPIE.5489...11K}}, first={Panoramic Survey Telescope \& Rapid Response System \citep[PanSTARRS;][]{2004SPIE.5489...11K}}}
\newglossaryentry{lsst}{name=LSST, description={Large Synoptic Survey Telescope}, first={Large Synoptic Survey Telescope \citep[LSST;][]{2006AAS...209.8604P}}}
\newglossaryentry{ppmxl}{name=PPMXL, description={PPMXL Catalog of Positions and Proper Motions on the ICRS \citep{2010AJ....139.2440R}}}
\newglossaryentry{gaia}{name=GAIA, description={Global Astrometric Interferometer for Astrophysics \citep{2001A&A...369..339P}}, first={Global Astrometric Interferometer for Astrophysics \citep[GAIA;][]{2001A&A...369..339P}}}
\newglossaryentry{ligo}{name=LIGO, description={Laser Interferometer Gravitational Wave Observatory}, first={Laser Interferometer Gravitational Wave Observatory \citep[LIGO;][]{1992Sci...256..325A}}}
\newglossaryentry{aligo}{name=Advanced LIGO, description={Advanced LIGO}, sort=ligo2}
\newglossaryentry{lisa}{name=LISA, description={Laser Interferometer Space Antenna \citep{1994ESAJ...18..219J}}, first={Laser Interferometer Space Antenna \citep[LISA;][]{1994ESAJ...18..219J}}}
\newglossaryentry{ccd}{name=CCD,description={Charged Coupled Device}, first={charged coupled device (CCD)}, firstplural={charged coupled devices (CCDs)}}

\newcommand{\sn}[2]{SN~#1#2}

\newcommand{\snr}[1]{SNR~#1}

\newglossaryentry{sn}{name=Supernova, text={SN}, plural={SNe}, description={exploding star}, nonumberlist=true}
\newglossaryentry{snia}{name=Type~Ia (SN~Ia), text={SN~Ia}, description={Thermonuclear explosion of a white dwarf - spectra show no hydrogen but a strong silicon line},first={Type~Ia supernova (SN~Ia)}, firstplural={Type Ia supernovae (SNe~Ia)}, plural={SNe~Ia}, parent=sn, nonumberlist=true}
\newcommand{\sneia}{\glspl*{snia}}
\newcommand{\snia}{\gls*{snia}}

\newglossaryentry{branchnormal}{name={branch-normal}, text=\textit{Branch-normal}, description={Large homogeneous class of Type Ia Supernovae, defined in \citet{1993AJ....106.2383B}}, first={\textit{Branch-normal} SNe Ia \citep{1993AJ....106.2383B}}, parent=snia} 
\newglossaryentry{91t}{name={91T-like}, description={Luminous class of Type Ia supernovae similar to \sn{1991}{T} \citep{1992AJ....103.1632P}} , first={91T-like}, parent=snia} 
\newglossaryentry{91bg}{name={91bg-like}, description={Faint class of Type Ia supernovae similar to \sn{1991}{bg} \citep{1992AJ....104.1543F}}, first={91bg-like}, parent=snia} 
\newglossaryentry{02cx}{name={02cx-like}, description={Peculiar class of Type Ia supernovae similar to \sn{2002}{cx} \citep{2003PASP..115..453L}}, first={02cx-like \sneia\ \citep{2003PASP..115..453L}}, parent=snia} 

\newglossaryentry{snibc}{name=Type~Ib/c, text={SN~Ib/c}, description={Collapse of the core of a massive star -  spectrum shows no hydrogen and no silicon line},first={Type~Ib/c supernova (SN~Ib/c)}, firstplural={Type~Ib/c supernovae (SNe~Ib/c)}, plural={SNe~Ib/c}, parent=sn}

\newglossaryentry{snib}{name=Type~Ib, text={SN~Ib}, description={Spectrum shows no hydrogen and no silicon, but helium line},first={Type Ib supernova (SN~Ib)}, firstplural={Type~Ib supernovae (SNe~Ib)}, plural={SNe~Ib}, parent=snibc}

\newglossaryentry{snic}{name=Type~Ic, text={SN~Ic}, description={Spectrum shows no hydrogen, no silicon and no helium line},first={Type~Ic supernova (SN~Ic)}, firstplural={Type~Ic supernovae (SNe~Ic)}, plural={SNe~Ic}, parent=snibc}

\newglossaryentry{snii}{name=Type~II, text={SN~II}, description={Collapse of the core of a massive star - spectrum shows strong hydrogen line},first={Type~II supernova (SN~II)}, firstplural={Type~II supernovae (SNe~II)}, plural={SNe~II}, parent=sn}

\newglossaryentry{sniib}{name=Type~IIb, text={SN~IIb}, description={Spectrum shows hydrogen and helium lines},first={Type~IIb supernova (SN~IIb)}, firstplural={Type~IIb supernovae (SNe~IIb)}, plural={SNe~IIb}, see=snib, parent=snii}

\newglossaryentry{sniip}{name=Type~II~Plateau (Type IIP), text={SN~IIP}, description={Lightcurve shows plateau},first={Type~IIP supernova (SN~IIP)}, firstplural={Type~II Plateau supernovae \citep[SNe~IIP;][]{1979A&A....72..287B}}, plural={SNe~IIP}, parent=snii}

\newglossaryentry{sniil}{name=SN~II~Linear, text={SN~IIL}, description={Lightcurve shows no plateau, but linear decline},first={Type~IIL supernova (SN~IIL)}, firstplural={Type~II~Linear supernovae \citep[SNe~IIL;][]{1990MNRAS.244..269S}}, plural={SNe~IIL}, parent=snii}

\newglossaryentry{sniin}{name=Type II narrow-lined (Type IIn), description={Spectrum shows narrow lines},first={Type~II~narrow-lined supernova (SN IIn)}, firstplural={Type~IIn supernovae (SNe~IIn)}, plural={SNe~IIn}, parent=snii}

\newglossaryentry{snr}{name=Remnant (SNR), text=SNR, description={Remnant left visible post-explosion}, first={supernova remnant (SNR)}, firstplural={supernova remnants (SNRs)}, parent=sn}

\newglossaryentry{dtd}{name=DTD,description={delay time distribution - expected supernova rate over time after a brief outburst of starformation},first={delay time distribution (DTD)}, firstplural={delay time distributions (DTDs)}, plural=DTDs}

\newglossaryentry{hvg}{name=HVG,description={high velocity gradient - Type Ia supernovae with a fast evolution of photospheric velocity},first={high velocity group (HVG)}, firstplural={high velocity groups (HVGs)}, plural=HVGs, parent=snia}

\newglossaryentry{lvg}{name=LVG,description={low velocity gradient - Type Ia supernovae with a slow evolution of photospheric velocity},first={low velocity group (LVG)}, firstplural={low velocity groups (LVGs)}, plural=LVGs, parent=snia}

\newglossaryentry{wd}{name=white dwarf (WD), text=WD, description={White Dwarf - extremely dense stellar remnant}, first={white dwarf (WD)}}
\newglossaryentry{onemgwd}{name= Oxygen/Neon (ONe), text={ONe-WD},description={Oxygen/Neon White Dwarf}, first={oxygen/neon White Dwarf (ONe-WD)}, parent=wd}
\newglossaryentry{cowd}{name=carbon/oxygen (CO), text={CO-WD}, description={carbon/oxygen white dwarf}, first={carbon/oxygen white dwarf (CO-WD)}, firstplural = {carbon/oxygen white dwarfs (CO-WDs)}, parent=wd}

\newglossaryentry{sds}{name=SD-Scenario,description={single-degenerate scenario (single white dwarf accreting from non-degenerate companion)}, first={single-degenerate scenario (SD-scenario)}}

\newglossaryentry{dds}{name=DD-Scenario, description={double degenerate scenario (merging of two white dwarfs)}, first={double-degenerate scenario (DD-scenario)}}

\newglossaryentry{sss}{name=SSS, text={supersoft \xray\ source}, description={supersoft \xray\ source - believed to be emitted by nuclear fusion on a white dwarf's surface}}

\newglossaryentry{amcvn}{name=AM CVn, description={AM Canum Venaticorum star (white dwarf accreting hydrogen poor matter from a companion star; see \cite{2005ASPC..330...27N})}}

\newglossaryentry{rlof}{name=RLOF, description={Roche Lobe Overflow (see \citet{1971ARA&A...9..183P} for a more detailed description)}, first={Roche-lobe overflow (RLOF)}}

\newglossaryentry{mchan}{name={Chandrasekhar mass~}, text={Chandrasekhar~mass}, symbol={\ensuremath{M_\textrm{Chan}}}, plural={Chandrasekhar~masses}, description={Mass when the core of a star collapses due to insufficient degeneracy pressure - for a white dwarf $\approx1.38\,M_\odot$ see \citet{1931ApJ....74...81C}}, first={Chandrasekhar~mass \citep[$M_\textrm{Chan}=1.38\,M_\odot$;][]{1931ApJ....74...81C}}, sort=mchan}

\newglossaryentry{w7}{name={W7 model},description={W7 model \citep{1984ApJ...286..644N}},first = {W7 model \citep{1984ApJ...286..644N}}}

\newglossaryentry{ew}{name=Equivalent Width, text={EW}, description={width of a rectangle that has the same area as a spectral line when taken to zero flux}, first={equivalent width (EW)}, firstplural={equivalent widths (EWs)}}
\newglossaryentry{agb}{name=AGB,description={Asymptotic Giant Branch}}
\newglossaryentry{cmb}{name=CMB,description={Cosmic Microwave Background}}
\newglossaryentry{csm}{name=CSM,description={Circumstellar Medium}, first={circumstellar medium (CSM)}}
\newglossaryentry{csi}{name=CSI,description={Circumstellar Interaction}, first={circumstellar interaction (CSI)}}
\newglossaryentry{ism}{name=ISM,description={Interstellar Medium}, first={interstellar medium (ISM)}}
\newglossaryentry{ige}{name=IGE,description={Iron Group Element}, first={iron group element (IGE)}, firstplural={iron group elements (IGEs)}}
\newglossaryentry{epm}{name=EPM,description={Expanding Photosphere Method \citep{1974ApJ...193...27K}}, first={Expanding Photosphere Method (EPM)}}
\newglossaryentry{aic}{name=AIC,description={Accretion Induced Collapse}, first={accretion induced collapse (AIC)}}
\newglossaryentry{ime}{name=IME,description={Intermediate Mass Element}, first={intermediate mass element (IME)}, firstplural={intermediate mass elements (IMEs)}}
\newglossaryentry{h0}{name=\ensuremath{H_0},description={Hubbles constant}}
\newglossaryentry{nse}{name=NSE,description={Nuclear Statistical Equilibrium}, first={nuclear statistical equilibrium (NSE)}}
\newglossaryentry{cdm}{name=CDM,description={Cold Dark Matter}}
\newglossaryentry{grb}{name=GRB,description={Gamma Ray Burst}, first={Gamma Ray Burst (GRB)}, firstplural={Gamma Ray Bursts (GRBs)}}
\newglossaryentry{donor}{name=donor,description={non-degenerate companion in the \gls{sds}}}
\newglossaryentry{mainsequence}{name=main sequence,description={main sequence star}}
\newglossaryentry{redgiant}{name=red giant,description={red giant star}}
\newglossaryentry{mlcs}{name=MLCS,description={Multicolor Light Curve Shape method \citep[MLCS;][]{1996ApJ...473...88R}}, first={Multicolor Light-Curve Shape method \citep[MLCS;][]{1996ApJ...473...88R}}}
\newglossaryentry{rsoph}{name=RS~Ophiuci ,description={white dwarf accreting from a red giant - assumed progenitor of the \gls{sds}}, sort=rsoph}
\newglossaryentry{usco}{name=U~Scorpii,description={white dwarf accreting from a main sequence star - assumed progenitor of the \gls{sds}}, sort=usco}
\newglossaryentry{rcw86}{name=RCW~86,description={supernova remnant sometimes associated with \sn{185}{}}, sort=rcw86}
\newglossaryentry{casa}{name=Cas~A,description={Cassiopeia A supernova remnant - probably a \gls{snib} event}}
\newglossaryentry{cepheid}{name=Cepheid,description={very luminous variable star with a strong luminosity period relationship}}
\newglossaryentry{urca}{name=Urca, text=\textit{Urca}, description={process predominatly contributing to cooling in stars. The \textit{Urca} process consists of alternating electron-capture and $\beta^{-}$ decay of two nuclei pairs.},sort=urca} 
\newglossaryentry{alphacen}{name=Alpha Centauri,description={one of the brightest stars in the night sky and a close binary}}
\newglossaryentry{pcygni}{name={P Cygni}, text={P Cygni},description={a hypergiant luminous blue variable with strong winds. Often referred to as a description for their line profiles showing a emission peak at the rest wavelength of the line and a blue-shifted absorption trough.}}

\newglossaryentry{teff}{name={effective temperature~}, text={effective temperature}, symbol={\ensuremath{T_\textrm{eff}}}, description={Temperature of a blackbody emitting the same total energy}, sort=teff}

\newglossaryentry{logg}{name={surface gravity~}, text={surface gravity}, symbol={\ensuremath{\textrm{log}\,g}}, description={gravity at the surface of a star}, sort=logg}
\newglossaryentry{feh}{name={metallicity~}, text={metallicity}, symbol=\textrm{[Fe/H]},description={iron abundance relative to the sun}, sort=feh}

\newglossaryentry{texp}{name={time since explosion~}, text={time since explosion}, text={time since explosion}, symbol={\ensuremath{t_{\rm exp}}},description={time since explosion (measured in days)}, sort=texp, first={time since explosion (\ensuremath{t_{\rm exp}})}}

\newglossaryentry{lmc}{name=LMC,description={Large Magellanic Cloud}, first={Large Magellanic Cloud (LMC)}, sort=lmc}
\newglossaryentry{smc}{name=SMC,description={Small Magellanic Cloud}, sort=smc}
\newglossaryentry{z}{name=\ensuremath{z},description={redshift}, sort=z}
\newcommand{\teff}{\glssymbol*{teff}}
\newcommand{\logg}{\glssymbol*{logg}}
\newcommand{\feh}{\glssymbol*{feh}}

\renewcommand{\sn}[2]{\object{SN~#1#2}}

\shortauthors{W.E. Kerzendorf et al.}
\shorttitle{SN1604 companion search}

\begin{document}

\title{A reconnaissance of the possible donor stars to the Kepler supernova}

\author{Wolfgang~E.~Kerzendorf\altaffilmark{1}\altaffilmark{2}, Michael Childress\altaffilmark{1}, Julia Scharw\"{a}chter\altaffilmark{3}, Tuan Do\altaffilmark{2}, Brian~P.~Schmidt\altaffilmark{1}} 
\email{wkerzend@mso.anu.edu.au}

\altaffiltext{1}{Research School of Astronomy and
Astrophysics, Mount Stromlo Observatory, Cotter Road, Weston Creek,
ACT 2611, Australia}

\altaffiltext{2}{Department of Astronomy and Astrophysics, University of Toronto, 50 Saint George Street, Toronto, ON M5S 3H4, Canada}

\altaffiltext{3}{Observatoire de Paris, LERMA (CNRS: UMR 8112), 61 Av. de l'Observatoire, 75014, Paris, France}

\begin{abstract}
The identity of Type Ia supernova progenitors remains a mystery, with various lines of evidence pointing towards either
accretion from a non-degenerate companion, or the rapid merger of two degenerate stars leading to the thermonuclear destruction of a white dwarf.
In this paper we spectroscopically scrutinize 24 of the brightest stars residing in the central $38\arcsec \times 38\arcsec$ of the SN 1604 (Kepler) supernova remnant  to search for a possible surviving companion star. We can rule out, with high certainty, a red giant companion star - a progenitor indicated by some models of the supernova remnant.  Furthermore, we find no star that exhibits properties uniquely consistent with those expected of a donor star down to $L>10\lsun $. While the 
distribution of star properties towards the remnant are consistent with unrelated stars, we identify the most promising candidates for further astrometric and spectroscopic follow-up. Such a program would either discover the donor star, or place strong limits on progenitor systems to luminosities with $L << \lsun$.
\end{abstract}

\maketitle
\section{Introduction}

\sneia\ provide a means to measure cosmological distances and are also major contributors to the chemical evolution of the Universe, enriching the cosmos with large amounts of iron-peak elements. Despite considerable effort to identify the systems that become SN Ia over the past decade, no consensus has yet emerged. It is widely accepted that \sneia\ are the thermonuclear explosions of a massive \gls{cowd} . The community suggests two main progenitor scenarios \cite[for a review see][and references therein]{2012NewAR..56..122W}. In the first scenario, the WD accretes mass from a non-degenerate companion (known as a donor star) until the central temperature and density exceed the cooling threshold and a run-away nuclear burning occurs. This is usually described as the \gls{sds}. In the second scenario, two WDs merge, leading to a cataclysmic explosion (\gls{dds}).

Both events make a number of predictions, however only the \gls{sds} offers a directly observable prediction - the remaining donor star. On the face of it, the \gls{sds} looks easily falsifiable. The donor star should remain \citep[e.g.][]{2000ApJS..128..615M} and be easy to detect in historical remnants. There has been considerable effort \citep{2004Natur.431.1069R, 2009ApJ...691....1G,2009ApJ...701.1665K,2012Natur.481..164S,2012Natur.489..533G,2012arXiv1210.2713K,2012ApJ...759....7K, 2012ApJ...747L..19E} to find the donor stars of ancient supernovae. Two remnants in the LMC \citep{2012ApJ...747L..19E, 2012Natur.481..164S} and \sn{1006}\ \citep{2012ApJ...759....7K, 2012Natur.489..533G} have proven to be void of progenitors to a relatively stringent detection limit. A possible progenitor has been found in SN 1572 \citep{2004Natur.431.1069R} although the validity of this connection has been challenged \citep{2012arXiv1210.2713K}. Despite, these negative results this does not necessarily disprove the validity of the \gls{sds}, as one can construct scenarios for the single-degenerate case, which produce surviving companions that do not stand out among their neighbouring stars \citep[e.g.][]{2012ApJ...760...21P, 2013arXiv1303.2691L, 2003astro.ph..3660P}. 

The recent discoveries of \gls{csm} around roughly a quarter of normal \sneia\ \citep{2007Sci...317..924P, 2009ApJ...702.1157S, 2011Sci...333..856S, 2012ApJ...752..101F} seemingly suggests two subtypes. As \gls{csm} interaction is favoured by the \gls{sds} and is rarer, the lack of direct detection of a donor star could just be the relative rarity of these objects.

In this work we turn to the last unexamined (in respect to donor stars) young Galactic supernova remnant - the remnant of \sn{1604}\ (also known as Kepler's SN). There's a dispute about the precise distance \citep[$4-6.4$\,\kpc; see][and references therein]{2012A&A...537A.139C}, and we choose to adopt the upper and more conservative limit of d=6.4~\kpc\ \citep{1999AJ....118..926R} for this work. Furthermore, there's a known maximum apparent brightness \citep[V=-3;][]{1971SvA....14..798P}, which, together with a usual absolute brightness of \sneia\  \citep[$M_V=-19.3$;][]{2011ApJ...732..129R} and an extinction of $A_V=2.8$ \citep{2007ApJ...668L.135R} leads to a distance of $\approx 5~\kpc$, consistent with the other measurements. Using the historic brightness provides an independent and direct measurement of distance modulus and extinction, which proves useful when testing the candidates' luminosities. Compared to its Galactic siblings (\sn{1006}\ and \sn{1572}) the Kepler \snr\ shows a very unusual structure and the presence of a nitrogen-rich shell and therefore had initially been identified as a core collapse \citep[e.g.][]{1987ApJ...319..885B}. However more recent studies affirm its \snia\ nature, due to - mainly - its large amount of Fe and little oxygen \citep{2007ApJ...668L.135R, 2012ApJ...756....6P}. \citet{2012A&A...537A.139C} attributes the shell to a progenitor with high mass loss prior to explosion - a calling card for the \gls{sds}. This large amount of interstellar matter suggests that Kepler might have belonged to the class of \sneia\ showing \gls{csm}\ interaction in their spectra. Finally, the remnant shows a high systemic velocity of  $250~\kms$ \citep{1991ApJ...366..484B,2003A&A...407..249S}. This high systemic velocity, including a radial velocity component of -180~\kms, should help make it possible to more easily identify a potential donor star, as any star associated with the SN should have -180~\kms\ of motion added to its orbital motion at the time of the explosion.

Here, we report a photometric and spectroscopic search for a donor star in the \sn{1604}\ remnant. In Section~\ref{sec:observations} we give an overview of the observation of the spectroscopic data and describe the \gls{hst} data. We describe the analysis of the spectroscopic data in Section~\ref{sec:analysis} and discuss its implications in Section~\ref{sec:discussion}. We conclude this paper and discuss possible future work in Section~\ref{sec:conclusion}.

\section{Observations}
\label{sec:observations}
\subsection{Integral Field Spectroscopy}
We used the \gls{ifu} called \gls{wifes} mounted on the ANU 2.3m Telescope at the Siding Spring Observatory  to undertake the spectroscopic observations in this study.. The \gls{wifes}-spectrograph is an image slicer with twenty five $38\times 1\arcsec$ slitlets and 0.5\arcsec\ sampling in the spatial direction on the detector. We chose this instrument for its large field of view ($25\arcsec\times38\arcsec$) and a resolution of $R=7000$ to provide the best possible radial velocity measurements. 

\begin{figure*}[tb!]
   \centering
   \includegraphics[width=\textwidth]{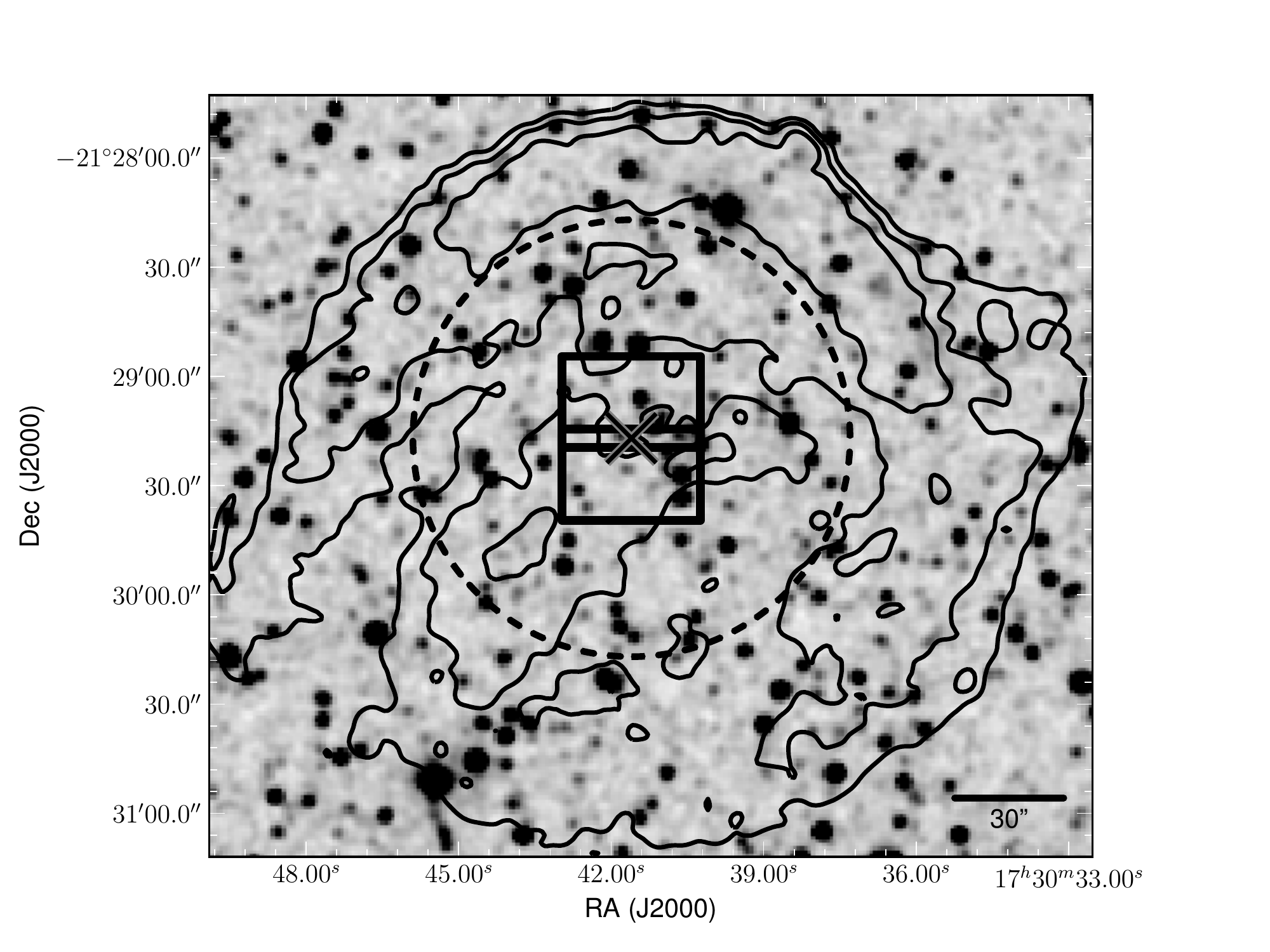} 
   \caption{SN1604 \gls{2mass} image with contours of ACIS X-Ray image \citep[ObsID 116; ][]{2002ApJ...581.1101H}. In this case, 19\arcsec\ equals 1420~\kms\ at 6.4~\kpc\ over 400 years. The dashed circle describes a 60\arcsec circle where we photometrically searched for potential companion stars.}
   \label{fig:sn1604_overlay}
\end{figure*}

We have adopted \rasc{17}{30}{41}{6} \decl{-21}{29}{17} as the centre of the remnant from \citet{2008ApJ...689..225K}. Due to the potential uncertainty in the centre determination \citep[for a more detailed discussion about the difficulty of centre determination see Section~2.2 in][]{2012ApJ...759....7K} we split up our observations to cover two fields of $25\arcsec\times38\arcsec$ with an overlap giving a total field of $38\arcsec\times38\arcsec$ (see Figure~\ref{fig:sn1604_overlay}). This setup gives us a minimum search distance of $19\arcsec$ from the proposed centre, corresponding to a velocity of $1420~\kms$ (in the plane of the sky over 400 years) assuming a remnant distance of $6.4~\kpc$ \citep{2012A&A...537A.139C}. 


We selected the I7000 grating which covers a spectral region  ($6830~-~9120~\AA$) with 0.2\AA (7~\kms) resolution. This region includes the \ion{Ca}{2} Triplet (8498\AA, 8542\AA, and 8662\AA) and maximises the Signal to Noise of the stars observed, which are heavily reddened by dust. The \ion{Ca}{2} provides good radial velocity measurements even in low \gls{signalnoise} ratio observations.

\begin{figure*}[tb!]
   \centering
   \includegraphics[width=.49\textwidth, bb = 0 0 500 400,clip]{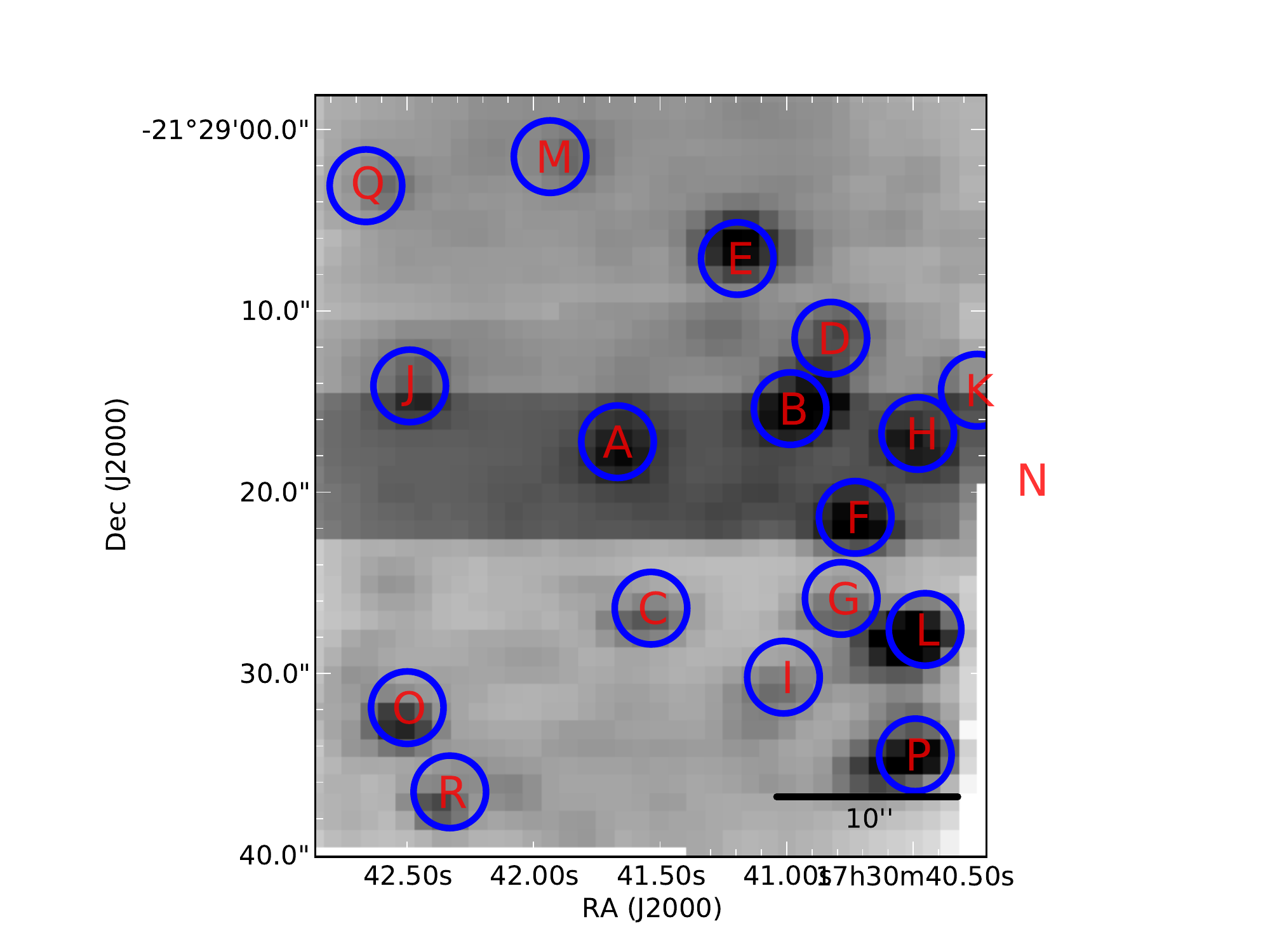} 
   \includegraphics[width=.49\textwidth, bb = 0 0 500 400,clip]{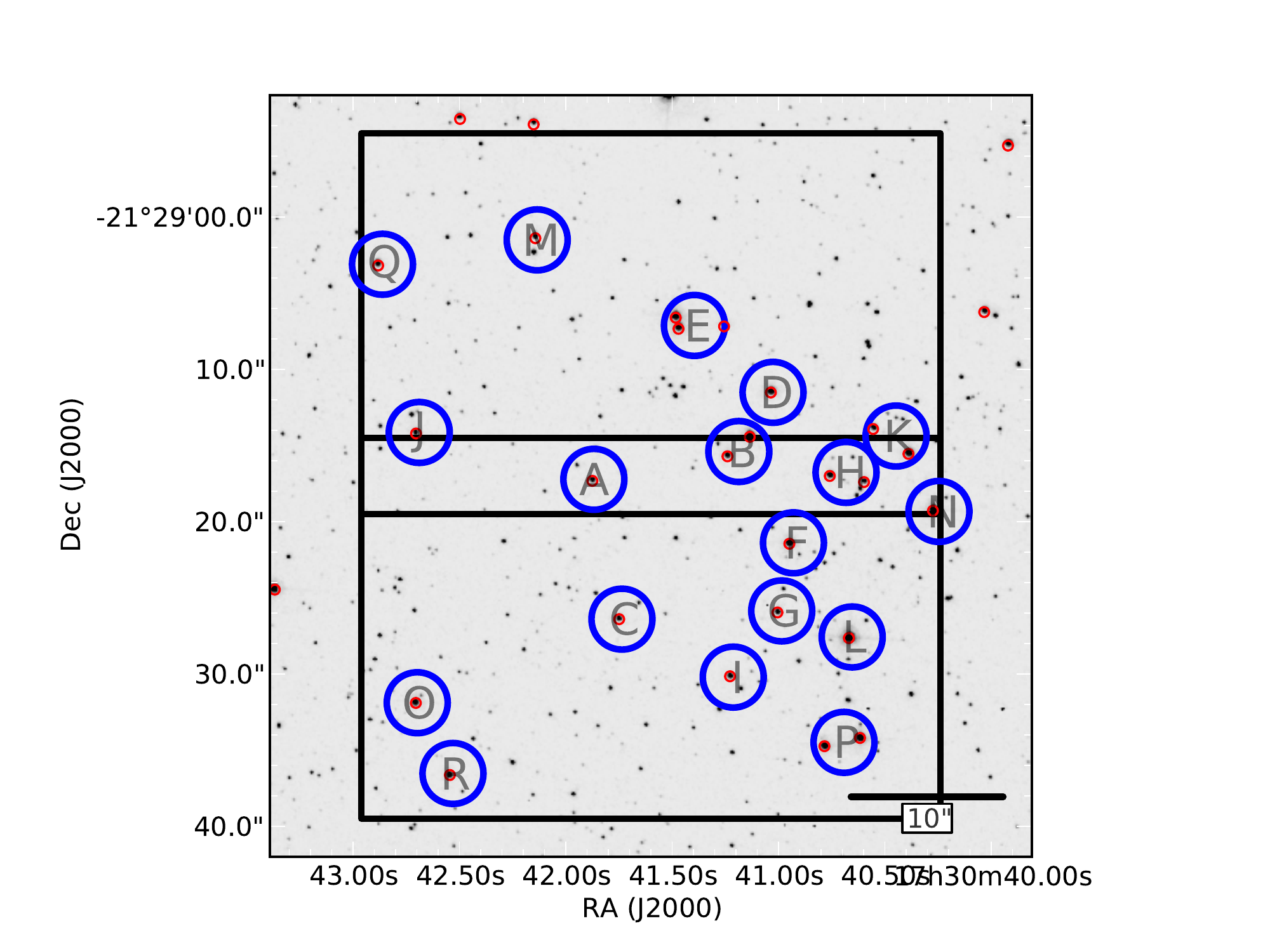}
   \caption{\textbf{Left panel:} The combination of a median (in the spectral direction) of two of our spectral cubes. The extraction regions marked in blue. The seeing is comparable to the 2MASS, enabling a direct comparison of 2MASS magnitudes and the spectra. Due to dithering, some of the stars described in the analysis (N and K) are not covered by the cubes displayed in this figure.\newline \textbf{Right~panel:} HST F550M image of the central of region of SN1604. We have marked the extraction regions with blue circles and the individual sources greater than $V=19.21$ ($10\lsun$ at the distance of the remnant taking the difference between F550M and V-filter into account) with red circles. There are a few extraction regions that contain more than one bright star .}
   \label{fig:sn1604_candidates}
\end{figure*}

The observations were carried out  on 2010 June 15 and 2010 June 16. The exposure time for all observations was $600~\textrm{s}$ with the southern field being observed 33 times ($5.5~h$) and the northern field observed 19 times ($3.2~h$) as weather conditions deteriorated. 
The seeing ranged between 1.7\arcsec and 2.7\arcsec with a median of 1.9\arcsec (see Figure~\ref{fig:sn1604_candidates} left panel for a sample of the data quality). 


WiFeS data were processed using the PyWiFeS package\footnote[1]{{\tt
http://www.mso.anu.edu.au/pywifes/}}.  Wavelength solutions for each
data cube were derived from night sky emission lines, achieving a
residual wavelength solution scatter of 0.1\AA.

In a final step, a simple spatial world coordinate system was applied to each cube with the help of \gls{2mass} for object coordinates (as \gls{2mass} shows a similarly broad PSF).

\subsection{Photometry}

To obtain accurate photometry, we used archival \gls{hst} data (\gls{hst} program GO-9731) observed in August 2003. The data consist of two F550M images with an exposure time of $240\,\textrm{s}$ each, observed with the \gls{acs}. We analyzed the frames using \gls{dolphot} and extracted magnitudes for all stars and corrected it to the V-Band (difference in flux between F550M and V for Vega is 0.12 magnitudes). Table~\ref{tab:hst_candidates} shows the magnitudes of stars with more than 10 solar luminosities in the V-Band at the distance of the remnant. In addition, we have supplemented the HST photometry of these objects with photometric information from the \gls{nomad}. Furthermore, Figure~\ref{fig:sn1604_candidates}  (right panel) shows the possible companion candidates and an overlay of the \gls{wifes}-fields onto the \gls{hst} image. 

\begin{deluxetable}{lcccccccccccc}
\rotate
\tablecaption{HST measurements of candidates \label{tab:hst_candidates}}
\tablehead{\colhead{Name} & \colhead{RA} & \colhead{Dec} & \colhead{F550M} & \colhead{Luminosity(V)\tablenotemark{a}}
& \colhead{NOMAD ID}\tablenotemark{b} & \colhead{B\tablenotemark{d}} & \colhead{V\tablenotemark{d}} & \colhead{R\tablenotemark{d}}
 & \colhead{J\tablenotemark{d}} & \colhead{H\tablenotemark{d}} & \colhead{K\tablenotemark{d}}\\
\colhead{-} & \colhead{hh:mm:ss.ss} & \colhead{dd:mm:ss.s} & \colhead{mag} & \colhead{$L_\odot$} &  \colhead{-}
& \colhead{mag} & \colhead{mag} & \colhead{mag} & \colhead{mag} & \colhead{mag} & \colhead{mag}}
\startdata
A1 & 17:30:41.88 & -21:29:17.3 & 18.93 & 9 & 0685-0474466 & -- & -- & 17.6 & 15.1 & 14.5 & 14.2\\ 
B1 & 17:30:41.13 & -21:29:14.4 & 17.55 & 33 & 0685-0474416\tablenotemark{b} & 18.0 & 16.7 & 13.9 & 14.0 & 13.2 & 12.9\\ 
B2 & 17:30:41.24 & -21:29:15.7 & 18.98 & 9 & 0685-0474416\tablenotemark{b} & 18.0 & 16.7 & 13.9 & 14.0 & 13.2 & 12.9\\ 
C1 & 17:30:41.75 & -21:29:26.4 & 18.90 & 9 & 0685-0474460 & -- & -- & 17.9 & 15.6 & 14.9 & 14.6\\ 
D1 & 17:30:41.04 & -21:29:11.5 & 18.07 & 20 & 0685-0474416 & 18.0 & 16.7 & 13.9 & 14.0 & 13.2 & 12.9\\ 
E1\tablenotemark{c} & 17:30:41.48 & -21:29:06.6 & 17.24 & 43 & 0685-0474437 & 17.1 & 16.1 & 15.3 & 13.6 & 12.9 & 12.6\\ 
E2\tablenotemark{c} & 17:30:41.47 & -21:29:07.3 & 17.79 & 26 & 0685-0474437 & 17.1 & 16.1 & 15.3 & 13.6 & 12.9 & 12.6\\ 
E3 & 17:30:41.26 & -21:29:07.2 & 18.95 & 9 & 0685-0474437 & 17.1 & 16.1 & 15.3 & 13.6 & 12.9 & 12.6\\ 
F1 & 17:30:40.95 & -21:29:21.5 & 17.95 & 23 & 0685-0474398 & 18.8 & -- & 15.9 & 14.4 & 13.7 & 13.4\\ 
G1 & 17:30:41.01 & -21:29:26.0 & 18.92 & 9 & 0685-0474409 & -- & -- & -- & 15.5 & 14.6 & 14.3\\ 
H1 & 17:30:40.76 & -21:29:17.0 & 18.69 & 11 & 0685-0474388 & -- & -- & -- & 15.0 & 14.2 & 14.0\\ 
H2 & 17:30:40.60 & -21:29:17.4 & 19.16 & 7 & 0685-0474388 & -- & -- & -- & 15.0 & 14.2 & 14.0\\ 
I1 & 17:30:41.23 & -21:29:30.1 & 19.18 & 7 & 0685-0474414 & 20.3 & -- & 17.1 & -- & -- & --\\ 
J1 & 17:30:42.70 & -21:29:14.2 & 19.14 & 8 & 0685-0474517 & 19.0 & -- & 17.8 & 15.4 & 14.8 & 14.4\\ 
K1\tablenotemark{c} & 17:30:40.39 & -21:29:15.6 & 17.91 & 23 & 0685-0474365 & -- & 16.4 & -- & 14.8 & 14.2 & 14.1\\ 
K2 & 17:30:40.56 & -21:29:13.9 & 19.24 & 7 & 0685-0474365 & -- & 16.4 & -- & 14.8 & 14.2 & 14.1\\ 
L1\tablenotemark{c} & 17:30:40.67 & -21:29:27.6 & 16.50 & 86 & 0685-0474384 & 17.4 & 15.9 & 12.8 & 13.1 & 12.3 & 12.0\\ 
M1 & 17:30:42.14 & -21:29:01.4 & 19.16 & 7 & 0685-0474479 & -- & -- & 18.0 & 15.8 & 15.2 & 14.8\\ 
N1\tablenotemark{c} & 17:30:40.27 & -21:29:19.3 & 17.38 & 38 & 0685-0474362 & 18.5 & 16.2 & -- & 14.0 & 13.2 & 13.0\\ 
O1 & 17:30:42.71 & -21:29:31.9 & 17.94 & 23 & 0685-0474522 & 18.7 & 17.7 & 16.7 & 14.7 & 13.9 & 13.6\\ 
P1 & 17:30:40.62 & -21:29:34.2 & 17.44 & 36 & 0685-0474383 & 18.1 & 16.3 & -- & 13.7 & 13.0 & 12.7\\ 
P2 & 17:30:40.78 & -21:29:34.7 & 17.89 & 24 & 0685-0474383 & 18.1 & 16.3 & -- & 13.7 & 13.0 & 12.7\\ 
Q1 & 17:30:42.88 & -21:29:03.2 & 18.55 & 13 & 0685-0474537 & -- & -- & 17.8 & 15.8 & 14.9 & 14.9\\ 
R1 & 17:30:42.54 & -21:29:36.6 & 18.77 & 11 & 0685-0474505 & -- & -- & 17.2 & 15.3 & 14.5 & 14.4\\ 

\enddata
\tablenotetext{a}{Assuming the candidate to be at the distance of SN1604 by using the distance modulus calculated from \sn{1604}'s apparent magnitude and average \sneia\ absolute magnitude we obtain $m_V - M_V=16.3$. If we, instead, assumed a distance of 6~kpc and an extinction of $A_V=2.8$
the luminosity increases by a factor of 1.4 ($m_V - M_V = 16.7$)}
\tablenotetext{b}{In addition to stars marked with the same letter (but different digit), the extraction Region B and D are also covered by one NOMAD ID}
\tablenotetext{c}{Candidate for follow-up ($L>20\lsun$ and $P(\textrm{donor}|\vrad)>1\%$)}
\tablenotetext{d}{Photometry from the NOMAD catalogue (accuracy $\approx 0.3~\textrm{mag}$)}
\end{deluxetable}

\section{Analysis}
\label{sec:analysis}

Using the HST images as a template we selected extraction regions in our spectral cubes with an aperture of 2\arcsec, a compromise between using the available signal and avoiding contamination from close stars. Despite the small aperture, some extraction regions contain two or three stars. The aperture was then summed at each spectral channel. A sky measurement was obtained in a similar fashion using an annulus with an inner radius of 4\arcsec and an outer radius of 6\arcsec. To alleviate contribution of neighbouring stars,  the individual pixels of the sky annuli were median combined during extraction. We scaled the skylines of the sky annuli to the sky lines of the spectra to avoid an over- or under-subtraction to best remove any residual background.

We concentrated on the wavelength region 8400--8700\AA\ for the next part of the analysis - as this contains the strong \ion{Ca}{2} lines which are useful for radial velocity measurements. The sky-frame scale was fit with a minimizer \citep{powell1964efficient} simultaneously with a continuum (third-order polynomial), and a model of the sun (\teff=5780, \logg=4.4, and \feh=0.0). Although the \gls{signalnoise} ratio is in principle high enough to measure stellar parameters, the uncertainty in the continuum placement due to sky
subtraction errors makes the parameter estimations unreliable. In addition, we lack accurate (0.02 mag) colour photometry (B, V, and R) of all the objects (some photometry is only available for a blend of objects) to aid in the determination of stellar parameters. The unreliable placement of the continuum does not affect the radial velocity measurements as they only rely on the position and not the depth of lines.

\begin{figure*}[tb!]
   \centering
   \includegraphics[width=\textwidth]{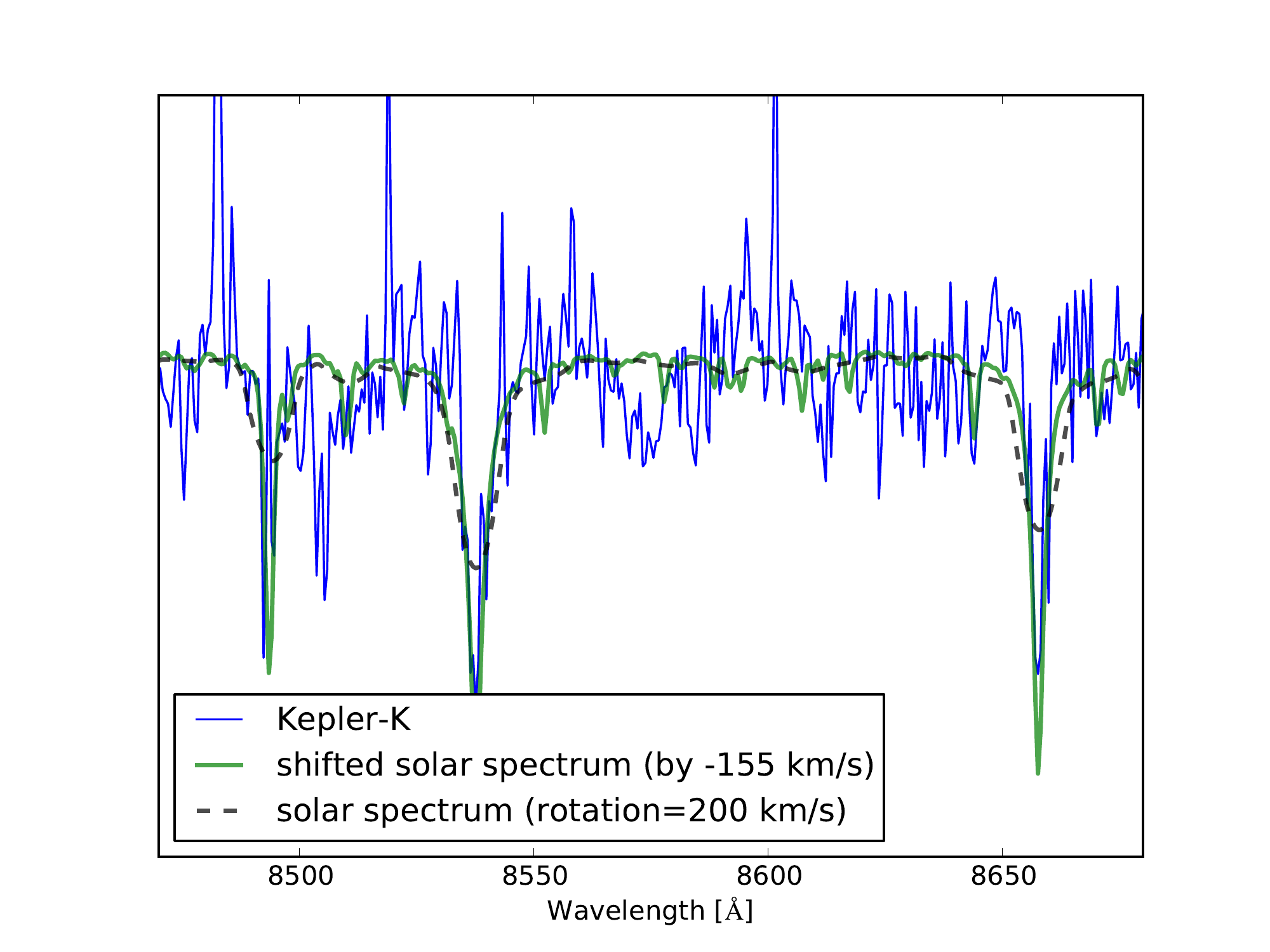} 

   \caption{The Kepler-K candidate showing a radial velocity of -155~\kms.  Such a velocity is consistent with that expected of more than a  third of the donor stars of the Kepler-SNR, but also of 5\% of unassociated stars in the direction of the remnant. Given that this study has analysed 20 stars' radial velocities, this velocity is not, on its own, indicative of an unusual star.}
   \label{fig:kepler-g}
\end{figure*}

The scaled sky was subtracted from the spectra and the resulting spectra (for each extraction site) were interpolated on a common wavelength grid and summed up. Next, we convolved a solar spectrum to the required resolution and shifted these synthetic spectra between -400\kms\ and 200\kms\ in a thousand equally spaced steps. We then compared the observed spectra with the set of synthetic spectra (using the root-mean-square technique) and chose the velocity corresponding to the lowest root-mean-square. Subsequently, we compared each fit with the spectrum and ascertained that the \ion{Ca}{2} features were clearly visible in the candidate spectrum and coincided with the shifted solar spectrum (see Figure~\ref{fig:kepler-g}). Furthermore, in the cases where HST photometry showed the presence of two stars, we checked the RMS fit for secondary minima that arises when there's two distinct sets of spectra. Only Kepler-H showed two clear sets of \ion{Ca}{2} features in the spectrum as well as in the RMS fit (henceforth designated as F1 and F2). Finally, we determined the typical error of the radial velocity to be $\approx4.5~\kms$  (or a tenth of a resolution element), by measuring the radial velocity in individual frames of a few candidates and looking at the resulting distribution. 

When compared to model spectra we find that the resolution and quality of the spectra does not allow for a reliable determination of the rotational velocity to better than 200~\kms\ (e.g. see Figure~\ref{fig:kepler-g}). As seen in Figure~\ref{fig:sn1604_candidates} we obtained spectra of sources down to $V = 19.2$ (which corresponds to $L_V = 10\lsun$ taking distance to the remnant and extinction into account). However, we could not obtain individual radial velocities for some of the sources that are blended (at R=7000 - corresponding to a resolution element width of 42~\kms\ - this is not unexpected when examining the radial velocity distribution in Figure~\ref{fig:sn1604_besancon}).


\begin{deluxetable}{lccc}
\tablecaption{Radial velocity measurements of candidates \label{tab:hst_vrad}\tablenotemark{a}}

\tablehead{\colhead{Name} & \colhead{\vrad} & \colhead{Distance from center} & \colhead{$P(\textrm{donor} | \vrad)$\tablenotemark{b}}\\
\colhead{Designation} & \colhead{\kms} & \colhead{arcsec} & \colhead{1}}
\startdata
A & -69.07 & 3.75 & 0.05 \\ 
B & 167.17 & 5.99 & 0.00 \\ 
C & 7.01 & 9.59 & 0.00 \\ 
D & -74.27 & 9.72 & 0.06 \\ 
E & -38.64 & 10.29 & 0.02 \\ 
F & -10.51 & 10.33 & 0.01 \\ 
G & -94.29 & 12.34 & 0.10 \\ 
H I\tablenotemark{c} & 177.58 & 12.81 & 0.00 \\ 
H II\tablenotemark{c} & 23.92 & 12.81 & 0.00 \\ 
J & 39.04 & 15.46 & 0.00 \\ 
K & -155.96 & 16.30 & 0.36 \\ 
L & -88.29 & 16.92 & 0.09 \\ 
M & 14.61 & 17.20 & 0.00 \\ 
N & -81.88 & 19.05 & 0.07 \\ 
O & -10.71 & 21.37 & 0.01 \\ 
P & 86.69 & 21.60 & 0.00 \\ 
Q & -59.06 & 22.43 & 0.04 \\ 
R & 41.44 & 23.45 & 0.00 \\ 

\enddata
\tablenotetext{a}{We could not reliably measure a radial velocity for Candidate I.}
\tablenotetext{b}{Probability to be the donor star for the given \vrad\ and assuming that 1 of 19 must be the donor star.}
\tablenotetext{c}{Extraction region H shows two different radial velocities, stemming from the two different stars visible in the HST image.}
\end{deluxetable}

\section{Discussion}
\label{sec:discussion}

\begin{figure*}[tb!]
   \centering
   \includegraphics[width=\textwidth]{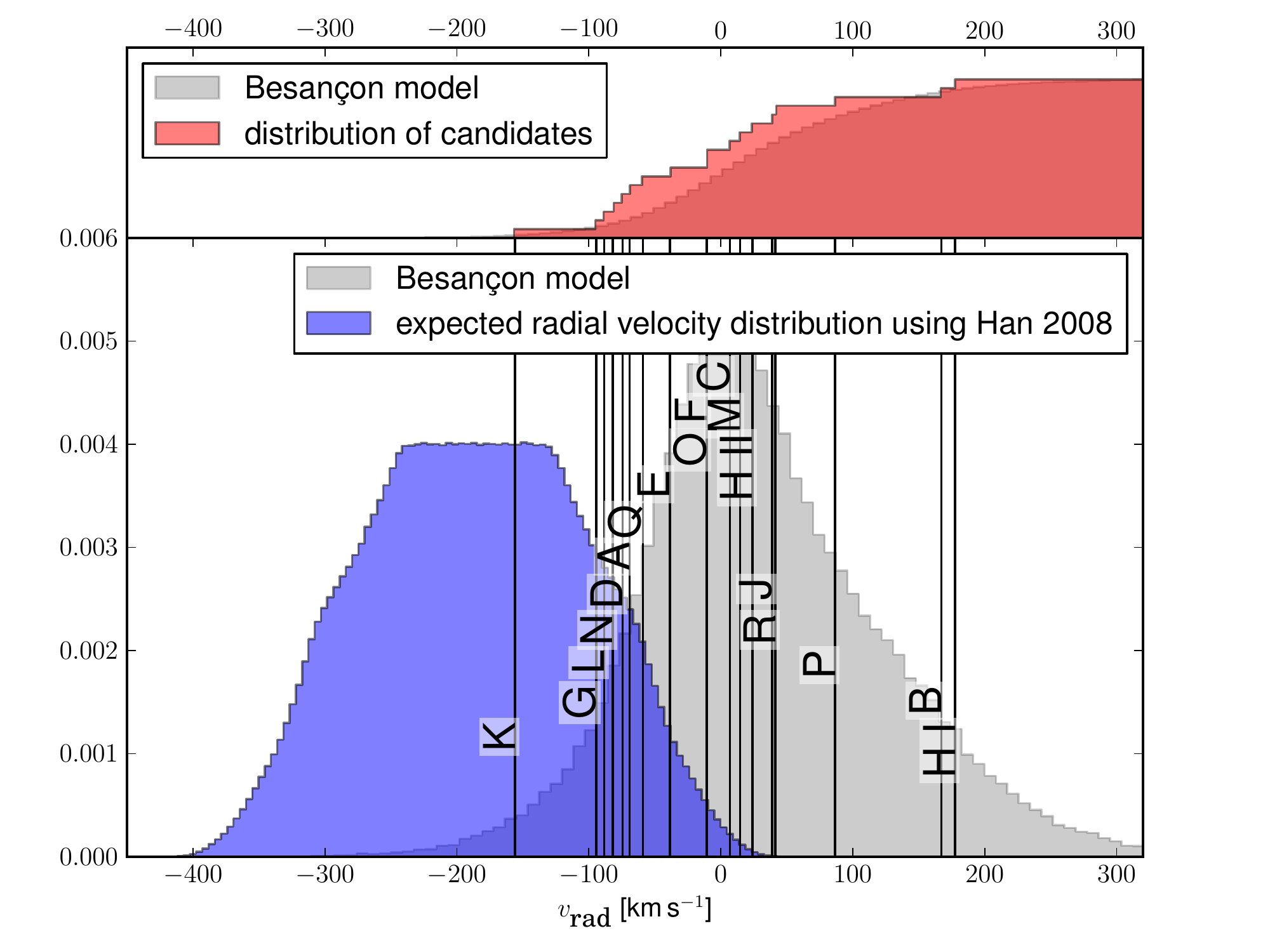} 

   \caption{Comparison of \gls{besancon} (1 square degree around Kepler, V cut between 15 and 20) and \citet{2008ApJ...677L.109H} and our measured values. The top panel shows the cumulative distribution for the \gls{besancon} and our measured values. The \citet{2008ApJ...677L.109H} distribution are the result of a Montecarlo simulation including the distribution of random angles and two \citet{2008ApJ...677L.109H} distributions}
   \label{fig:sn1604_besancon}
\end{figure*}

The \gls{sds} predicts an escape velocity of the donor of up to $200~\kms$ for main sequence stars and less for sub giants or giant stars \citep[down to roughly 60~\,\kms;][]{2008ApJ...677L.109H}. In the case of \sn{1604}, this velocity signature might very easily be lost in the kinematic signature of the Galaxy, except that the systemic velocity of the Kepler remnant helps separate candidates from the normal motions of stars along the line of sight. We compare our radial-velocity measurements to theoretical predictions by \citet{2008ApJ...677L.109H} and the \gls{besancon} of galactic dynamics (choosing 1 square degree area around \sn{1604}\ and limiting to magnitudes between $15<$V$<20$). In Figure~\ref{fig:sn1604_besancon} we use a Monte Carlo simulation using the distribution of radial velocities from \citet{2008ApJ...677L.109H} (including main-sequence to giant donors) and a random distribution of ejection angles to obtain a radial velocity distribution. Furthermore, we center this on \snr{1604}'s systemic radial velocity of $-185~\kms$\ \citep{2003A&A...407..249S}. When looking at the \citet{2008ApJ...677L.109H} distribution in Figure~\ref{fig:sn1604_besancon}, half of the possible cases would produce a star being a significant outlier compared to the unrelated background and foreground stars. However, none of our stars are significant outliers to the \gls{besancon}, with our measured radial velocity distribution agreeing very well with the \gls{besancon}.

We have performed a Bayesian analysis in Table~\ref{tab:hst_vrad} using,

\begin{align}
P(\textrm{donor}|\vrad) = P(\textrm{donor})P(\vrad;\textrm{Han2008})\nonumber\\
 / (P(\textrm{donor})P(\vrad;\textrm{Han2008})\nonumber \\
+ P(\textrm{not donor})P(\vrad;\textrm{\gls{besancon}}))\nonumber,
\end{align}

where $P(\textrm{donor}) = \frac{1}{20}$ as we assume one star to be the donor star and all other probabilities are determined from the probability distributions shown in Figure~\ref{fig:sn1604_besancon}. Kepler-K can be seen as the most anomalous star studied but using Bayesian analysis we found that finding an unrelated star that shows a velocity as low or lower than  Kepler-K is not unusual given our sample size of 20 stars and the given \gls{besancon} distribution. However, without additional information, Kepler-K's velocity does not stand-out from the expected velocity distribution for stars along the line of site to Kepler. If we add a prior that one of the 19 stars is associated with the remnant, then Kepler-K is the most likely star to be associated (36\% chance - or approximately 4 times more likely than the next most likely star, Kepler-G).

\citet{2000ApJS..128..615M} suggests that giant donor stars will be on the order of 1000~\lsun\  for at least 100,000 years post-explosion. None of the stars in our sample are in that brightness range (the brightest star being Kepler-L with $86~\lsun(V)$ at 6.4~\kpc). Therefore, there seems to be no viable giant donor located in \sn{1604}. This is evidence against the AGB model suggested by \citet{2012A&A...537A.139C} and \citet{2013ApJ...764...63B}, which is one scenario explaining the unusual \gls{csm} that surround the Kepler remnant, with a $4 - 5 \msun$ AGB donor. \citet{2012A&A...537A.139C} suggest a star with $m_v=12.0 \pm 0.5$ which is not present in our candidates. 

\cite{2013arXiv1305.0567W} have suggested that the centre of remnants is not easily determined and suggest widening the progenitor searches. We have used the \gls{nomad} catalog to find bright companions in a 60\arcsec circle around our current centre (see Figure~\ref{fig:sn1604_overlay}). The brightest star in K-Band reaches K=10.5 (43\arcsec from the center)  the brightest star in V-Band reaches V=14.7 or $L_V=330~\lsun$ (40\arcsec from the center).   Finally, \citet{2012A&A...537A.139C}  suggest the possibility of losing the entire envelope and only having a white dwarf left over ($0.8~\msun$).  A normal white dwarf would be below our detection limit. However, \citet{2000ApJS..128..615M} suggest that losing the entire envelope would not result in a normal white dwarf, but a much brighter and hotter object ($\approx 10^3 \lsun$).

For main-sequence and sub-giant companions two new studies \citep{2012arXiv1205.5028S,2012ApJ...760...21P} suggest a luminosity of $\approx 20 - 275 \lsun$ for companion stars 400 years after the explosion \citep[see Figure~6 in ][]{2012ApJ...760...21P}. Restricting ourselves to objects that have $\ge 1 \%$ probability and $\lsun > 20$, we are left with five candidates (E1, E2, K1, L1, and N1). These are the most notable stars for further follow-up. In addition, a run-away star can be hidden in the general distribution of radial velocities in the direction of \sn{1604}. This is due to both, the very broad distribution of radial velocities (as seen in Figure~\ref{fig:sn1604_besancon}) and the possibility of a high proper motion coupled with a low radial velocity. Obtaining new HST-ACS images would help isolate any associated star by utilizing all three velocity components, and enable candidates to be selected to much lower luminosities given the depth of the HST photometry.. 

Another interpretation is that there are no main-sequence or sub-giant donors in \sn{1604}. But there are some caveats to any such statement.  In some cases we extract the spectrum of two or more stars due to their close proximity on the sky, but do not find two radial velocities (except in extraction aperture Kepler-H) , meaning that a star's velocity is missing or that the two radial velocities are indistinguishably close. Possible stars with $L>20~\lsun$ include stars E2 and P2, although with their relative brightnesses compared to the primary star, a large negative radial velocity, consistent with a donor star, should have been visible.


\section{Conclusion \& Future Work}

\label{sec:conclusion}

In this work, we present photometric and spectroscopic observations of candidate stars in the centre of the \sn{1604}\ remnant. We can rule out red giant companions due to the lack of bright stars in the field (brightest star Kepler-L with V=16.5). In addition, we can rule out many candidates in the field, that have inconsistent radial velocity signatures (all stars with a $\vrad>0~\kms$ have $< 1\%$ probability of being associated with the explosion). Finally, our radial velocity measurements do not identify any peculiar star in this data set beyond what is expected from a Galactic distribution, and are therefore consistent with \snia\ progenitor scenarios which do not leave behind a bright donor star. This finding is similar to work on other historical remnants such as those which have scrutinized \sn{1006}\ \citep{2012Natur.489..533G,2012ApJ...759....7K}, and the LMC results \citep{2012ApJ...747L..19E, 2012Natur.481..164S}. 

Both the newly discovered \gls{csm} interacting \sneia\ and the lack of donor stars requires an explanation that neither the traditional \gls{sds} or the traditional \gls{dds} can provide. \citet{2011ApJ...730L..34J,2011ApJ...738L...1D,2012ApJ...744...69H,2012ApJ...756L...4H} have suggested a modified \gls{sds} in which the companion has time to evolve and become a white dwarf before the explosion of the primary or is intrinsically dim \citep{2012ApJ...758..123W}. This can explain the difficulty in finding a stellar companion. If a \snia\ is the result of two white dwarfs merging in a common envelope, this evolution might explain the \gls{csm} interaction that is normally not expected for a \gls{dds} - it also explains the lack of a companion \citep[][van Kerkwijk priv. comm. ]{2011MNRAS.417.1466K}.  

To further test main-sequence and sub-giant progenitor scenarios, additional observations are called for. The current spectroscopic observations only provide an upper limit of $\vrot=200~\kms$, which is insufficient to discriminate between unassociated sources with typically not more than a few 10~\kms\ and donor stars with 60-140~\kms \citep[see Figure~4 in ][]{2012ApJ...759....7K}. We propose high-resolution spectra of the viable stars with $L>10\lsun$ (E1, E2, K1, L1, and N1) with a multi-object fibre system such as VLT+FLAMES. Such observations would be able to both detect rotation and better detect the blended stars kinematic signatures. In addition, a single ACS HST follow-up observation would allow an accurate measurement of all of the remnant's stars' proper motions, enabling detection of objects with motions consistent with the remnant down to the luminosities expected for even a hot white dwarf. 

Currently only three remnants have been spectroscopically searched for donor stars (SN1572, SN1006 and SN1604), which is a sample plagued by small number statistics. The fourth known Milky Way Type Ia \snr~RCW86 also lends itself for a donor star search. At first glance, this remnant with its lower extinction \citep[$A_V \approx 1.7$;][]{1983MNRAS.204..273L} and close proximity to Earth \citep[$d=2.5~\kpc$; ][]{2011ApJ...741...96W}) makes it seem ideal. However, there are two seemingly separate expansion fronts (one in the south west and one in the north east) that would make a center determination very difficult. A safe approach is a search area that encompasses both centers, however this would involve scrutinizing roughly 2000 stars to go down to a luminosity of $L=10\lsun$. This makes RCW86 currently not feasible for a search and thus Kepler is the last remnant easily searched in the Galaxy. In the future, GAIA will measure parallax distances and proper motions, which will enable us to narrow down the possible candidate stars for RCW86.

\section{Acknowledgements}

We like to thank Ben Shappee for useful discussions on post explosion donor evolution. Furthermore, Steve Reynolds, Kazik Borkowski, Mary Burkey, and Jacco Vink were providing crucial points about a possible AGB star donors. In addition, we would like to thank Carles Badenes for his description of the literature pertaining to the identity of Kepler's SN. Finally, we would like to thank the anonymous referee for helpful and constructive comments. 

J.S. acknowledges the European Research Council
for the Advanced Grant Program Num 267399-Momentum.

\bibliographystyle{hapj}
\bibliography{sn1604_wifes}

\end{document}